\documentclass{pasj00}
\onecolumn
\begin{document}
\SetRunningHead{A. Kawata et al.}
{How Do We See the Relativistic Accretion Disk during Thermal Instability?}
\Received{2005/11/14}
\Accepted{????/??/??}

\title{How Do We See the Relativistic Accretion Disk during Thermal Instability?}

\author{Akihiro \textsc{Kawata},\altaffilmark{1} %
 Ken-ya \textsc{Watarai},\altaffilmark{1,2} and Jun \textsc{Fukue}\altaffilmark{1} }

\altaffiltext{1}{Astronomical Institute, Osaka Kyoiku University, Asahigaoka, Kashiwara, Osaka 582-8582}
\altaffiltext{2}{Research Fellow of the Japan Society for the Promotion of Science}
\email{j049340@ex.osaka-kyoiku.ac.jp}


\maketitle

\begin{abstract}
We calculated the bolometric images of relativistic slim disks
 during radiation-pressure-driven thermal instability. 
When the mass-accretion rate exceeds the critical one, 
 an inner region of the standard accretion disk bursts to
 change to a slim disk state having a large scale height. 
That is, the inner region of the disk becomes high temperature,
 and the thickness of the disk increases 
due to the increase of the radiation pressure. 
As a result, we found that the observed image of the disk during the burst strongly depends on the inclination angle.  
That is,
 radiation from the innermost disk would be occulted by
 the disk outer rim for high inclination angles
 ($i~\gtrsim~70^\circ$). 
We also calculated the spectral energy distribution
 during thermal instability. 
The Wien peak of the spectrum of high inclination angles becomes softer
 than that of low inclination angles due to the geometrical thickness. 
From these facts, even if the burst happens in black hole candidates, we may not observe the burst when the inclination angle is large.
We may suggest that numerous luminous black hole candidates 
 are still hidden in our Galaxy. 
\end{abstract}

\section{Introduction}
Observational features of microquasars are considerably similar to those of quasars. 
That is, both microquasars and quasars have relatively high luminosities,
 relativistic jets, and central black holes
 (Mirabel, Rodr\'{\i}guez 1999; Fender, Belloni 2004 for recent reviews). 
Apart from the difference of black hole masses,
 an investigation of microquasars is strongly linked to
 understandings of quasar activities. 
Hence, many astronomers are engaged in studying this field. 

The galactic object GRS~1915+105 is 
one of the famous microquasars in our Galaxy
 owning to the unique observational behavior;
 i.e., the speed of jets is about $0.92~c$, quasi-periodic,
 flare like bursts, and limit-cyclic luminosity variations
 (Mirabel, Rodr\'{\i}guez 1999; Belloni et al. 2000; Fender, Belloni 2004). 
Although curious characteristics in this object collect attention of many astronomers,
the pattern of variability is
 too complicated to understand well (Belloni et al. 2000). 
The X-ray light curves of GRS 1915+105 in its high luminosity state
 show limit-cyclic bursting behavior
 on the timescale of a few tens of seconds. 
The typical timescale of the limit-cyclic phenomena
 corresponds to a viscous time scale in an accretion disk. 
Accordingly, the phenomenon is attributed to the radiation-pressure-driven
 thermal instability around the disk inner region (Belloni et al. 1997a, 1997b). 

The theory of the radiation-pressure-driven thermal instability
 in accretion disks was investigated by many authors
 (Shibazaki, H\=oshi 1975; Shakura, Sunyaev 1976). 
Honma et al. (1991) was the first to numerically
 demonstrate the limit-cycle oscillations
 between the two different states, based on a time-dependent calculation. 
Unfortunately, this kind of instability could occur only
 when the mass-accretion rate is larger in comparison with a critical one,
 $\sim 0.1 L_{\rm E}$, and a candidate where such an instability
 occurs was not found until microquasars were discovered. 
They show that an existence of an adequately radiation-pressure dominated
 region is essential for developing thermal instability.
After the discovery of microquasar GRS 1915+105, 
Szuszkiewicz and Miller (1997, 1998)
 confirmed the results of Honma et al. (1991), and moreover,
 they showed the difference of viscosity parameters, $\alpha$,
 during thermal instability. 
The emergent spectrum during thermal instability was studied
 by Zampiri et al. (2001). 
Watarai and Mineshige (2003a; hereafter WM03)
 explained the limit-cycle variations of the maximum temperature
 and the emitting radius in GRS~1915+105. 
Recentry, it is confirmed that 2D SPH calculations also reproduce the limit-cycle oscillation (Teresi et al. 2004a,b)
When we compare an observation with a disk model,
 there remain two important elements to examine more carefully. 
They are ``geometrical thickness of the disk'' and
 ``special/general relativistic effects''. 
If the accretion rate increases, the thickness of the disk rapidly increases
 due to the radiation pressure within the disk (Shakura, Sunyaev 1973).  
In addition, the ``inner edge'' of the disk in the supercritical
 accretion decreases down to less 
than 3 $r_{\rm g}$ (Watarai, Mineshige 2003b). 
Hence, we should take into account special/general relativistic effects;
 i.e., transverse Doppler effect, gravitational redshift,
 and light bending via curvature of space-time. 
These two important elements are strongly related to
 a viewing angle from a disk to a distant observer. 
Dependence of inclination angles in different accretion rates
 was pointed out by Watarai et al. (2005),
 and the observed spectrum significantly changes
 with high inclination angles and high mass accretion rates.
However, they do not consider in the time evolution of the disk. 

Accordingly, we tried to examine the observational spectra
 during thermal instability in WM03
 with various inclination angles. 
Note that previous papers have never considered the geometrical
 and special/general relativistic effects simultaneously
 during thermal instability. 
Thus, we examine there two effects in this paper.

In the next section, 
 we briefly overview the mechanism of thermal instability
 and their characteristics using the results of WM03. 
We show the flux images and the X-ray spectra of
 the inner disk during thermal instability in section 3. 
The comparison with the X-ray observation of the microquasar
 GRS~1915+105 is discussed in section 4. 
The final section contains our concluding remarks. 


\section{Disk Structure during Thermal Instability}
\subsection{Basic Equations}
The basic equations for the viscous accretion disk
are shown in the textbook by, e.g., Kato et al. (1998).
We here briefly summarize the basic equations.

We use the cylindrical coordinates ($r$, $\varphi$, $z$),
where the $z$-axis is the symmetrical one of the disk. 
We adopt a pseudo-Newtonian potential, $\psi = -GM/(R-r_{\rm g})$ with
$R \equiv \sqrt{r^{2}+z^{2}}$ as the effect of general relativity (Paczynsky \& Wiita 1980),
where $r_{\rm g}$ is the Schwarzschild radius 
defined by $r_{\rm g} = 2GM/c^{2} = 3\times10^6(M/M_{\odot})$ cm.
The disk is assumed to be axisymmetric.
The effects of outflow/jet or coronal dissipation are
 ignored in this calculation.

We adopt one-zone approximation in the vertical ($z$) direction, 
and the physical quantities are integrated in the vertical direction;
i.e., the surface density is $\Sigma = \int_{-H}^{H} {\rho} dz = 2I_{\rm N} \rho H$, 
and the height-integrated pressure is $\Pi = \int_{-H}^{H} {p} dz = 2I_{\rm N+1} p H$. 
Here,  $\rho$, $p$ and $H$ are the mass density, the total pressure,
 and the scale height, respectively.
 The coefficients $I_{\rm N}$ and $I_{\rm N+1}$ were introduced 
by H\=oshi (1977).
In the vertical direction,
the density and pressure are assumed to be
related by the polytropic relation
 $p \propto \rho^{1+1/N}$.
We assign $N=3$ in all calculations ($I_3 = 16/35$, $I_4=128/315$). 
The height-integrated equation of state is
\begin{equation}
\Pi = \Pi_{\rm rad} + \Pi_{\rm gas} 
=  \frac{a}{3} T_{\rm c}^{4} \cdot 2 H 
+ \frac{k_{\rm B}}{\bar{\mu}m_{\rm H}} \Sigma T_{\rm c},
\label{eq:pres}
\end{equation}
where the first term on the right-hand side represents the radiation pressure 
($a$ and $T_{c}$ are the radiation constant and the temperature on the equatorial plane, respectively)
 and the second term represents the gas pressure
 ($k_{\rm B}$ is the Boltzmann constant, $\bar{\mu}$ = 0.5 is the mean molecular weight,
 and $m_{\rm H}$ is the hydrogen mass).
We assume that, during thermal instability,
the hydrostatic equilibrium holds in the vertical direction of the disk,
\begin{equation}
(2N + 3) \frac{\Pi}{\Sigma} = H^2\Omega_{\rm K}^2,
\label{eq:hydro}
\end{equation}
where $\Omega_{\rm K}$ denotes the Keplerian angular frequency 
under the pseudo-Newtonian potential.

The mass conservation, the radial component of the momentum conservation, 
and the angular momentum conservation are respectively written as follow;
\begin{eqnarray}
\frac{\partial}{\partial t} (r \Sigma) 
+ \frac{\partial}{\partial r} (r \Sigma v_{r}) &=& 0, 
\label{eq:cont}
\\
\frac{\partial}{\partial t} (r \Sigma v_{r})
 + \frac{\partial}{\partial r} (r \Sigma v_{r}^2 + r \Pi)
 &=& \Sigma (v_{\varphi}^2 - v_{\rm K}^2)       
 + \left(1- \frac{d \ln \Omega_{K}}{d \ln r} \right) \Pi,
\label{eq:rmom}
\\
\frac{\partial}{\partial t} (r^2 \Sigma v_{\varphi}) 
 + \frac{\partial}{\partial r} (r^2 \Sigma v_{r} v_{\varphi} - r^2 T_{r \varphi}) 
 &=& 0,
\label{eq:pmom}
\end{eqnarray}
where $v_{r}$ and $v_{\varphi}$ are the radial and azimuthal components
of velocity, respectively,
and related to the angular momenta 
by $l \equiv r v_{\varphi}$ and 
$l_{\rm K} \equiv r v_{\rm K} = r^2 \Omega_{\rm K}$, respectively.

The $r \varphi$-component of viscous stress tensor in equation (\ref{eq:pmom}) is prescribed as
\begin{equation}
 T_{r \varphi} = \int^{H}_{-H} {t_{r \varphi}} dz 
  = - \alpha \beta^{\mu} \Pi
\label{eq:vis}
\end{equation}
where $\mu$ is a parameter ($0 \lesssim \mu \lesssim 1$) and 
$\beta$ $\equiv$ $p_{\rm gas} / p$.
In the limit of $\mu \rightarrow 0$, we recover the ordinary 
$\alpha$ viscosity prescription (Shakura \& Sunyaev 1973).
If we take the limit of $\mu \rightarrow 1$,
 the viscosity depends on the gas pressure only.

Finally, the energy equation is
\begin{equation}
\frac{\partial}{\partial t} (r \Sigma \epsilon_{\rm tot})
 + \frac{\partial}{\partial r} 
\left[
r (\Sigma \epsilon_{\rm tot} + \Pi) v_{r} - r T_{r \varphi} v_{\rm \varphi}
\right ]
   = -2rF^{-}
\label{eq:ene}
\end{equation}
in which advective cooling, viscous heating, 
and radiative cooling are considered.
The explicit form of the total energy $\epsilon_{\rm tot}$ is
\begin{equation}
\epsilon_{\rm tot} =
 \left[ 3( 1 - \beta) + \frac{\beta}{\gamma-1} + \frac{1}{2} \right]
  \frac{\Pi}{\Sigma}
   + \frac{1}{2}(v_{r}^2 + v_{\rm \varphi}^2) + \psi_0(r)
\label{eq:etot}
\end{equation}
where the first term on the right hand side is the internal energy of the gas
($\gamma$ is the adiabatic index and 
we set $\gamma = 5/3$ in the present calculation),
the second term represents the kinetic energy, 
and the last term is the potential energy on the equatorial plane,
$\psi_0 \equiv -GM/(r -r_{\rm g})$.
Radiative cooling flux per unit surface area in an optically thick medium
 is given by
\begin{equation}
F^{-} = \frac{8acT_{c}^4}{3 \bar{\kappa} \Sigma}
\label{eq:rad}
\end{equation}
where $\bar{\kappa}$ is the average opacity written
 by an optical depth $\tau$ as follow,
\begin{equation}
\tau = \bar{\kappa} \Sigma = (\kappa_{\rm es} + \kappa_{\rm ff}) \Sigma
\label{eq:opacity}
\end{equation}
 where $\kappa_{\rm es}$ = 0.4 is the opacity of the electron scattering,
 $\kappa_{\rm ff}$ = 0.64 $\times$ $10^{23}$ $\bar{\rho} \bar{T}^{-7/2}$
 is the absorption opacity via thermal Bremsstrahlung,
 and $\bar{\rho} =(16/35) \rho$ and $\bar{T} =(2/3) T_{\rm c}$ are
 vertically average density and temperature, respectively.

We solve equation (\ref{eq:cont})--(\ref{eq:ene}) 
 by the modified Lax Wendroff method with artificial viscosity.
The calculations are performed from the outer radius at 2000 $r_{\rm g}$
 down to the inner radius at $\sim 2.2 r_{\rm g}$. 
The total mesh number is 250.
We fixed black hole mass to be $m$ = 10
 ($m = M / M_{\odot}$, where $M_{\odot}$ is solar mass).
The critical accretion rate, $\dot{M}_{\rm crit}$, 
is defined by $L_{\rm E} / c^2$,
 where $L_{\rm E}$ is the Eddington luminosity and $c$ is the speed of light.
We define the dimensionless accretion rate to be
 $\dot{m} \equiv \dot{M}/\dot{M}_{\rm crit} = \dot{M} c^2 / L_{\rm E}$
 throughout the present study.

\subsection{What Is Thermal Instability?}
In this section, we briefly review the disk structures during thermal instability. 
Generally speaking,
 thermal instability in an accretion disk occurs
 in radiation-pressure-dominated regions (Shibazaki, H\=oshi 1975; Shakura $\&$ Sunyaev 1976).
 
In standard accretion disk regimes, where the mass-accretion rate is lower than the critical one,
 the viscous heating energy, $Q_{\rm vis}^+$, is first transformed
 to thermal energy,  
and finally released to the energy as a form of radiation $Q_{\rm rad}^-$.
That is, the accretion time of the disk gas is longer than
 the radiative cooling time,
 thus the disk keeps the equilibrium state (Shakura, Sunyaev 1973). 

However, when the mass-accretion rate at the disk inner region increases, 
 the radiative cooling energy cannot balance with the viscous heating energy. 
This is because that the dependence of a mass-accretion rate of the radiative cooling
 is $Q_{\rm rad}^{-} \propto \dot{M}^{1/2}$,
 while $Q_{\rm vis}^{+} \propto \dot{M}$ for the viscous heating.
When the $\dot{M}$ exceeds the critical rate,
no equilibrium solution exists (Kato et al. 1998). 

\begin{figure}[htb]
  \begin{center}
\hbox{
    \FigureFile(75mm,75mm){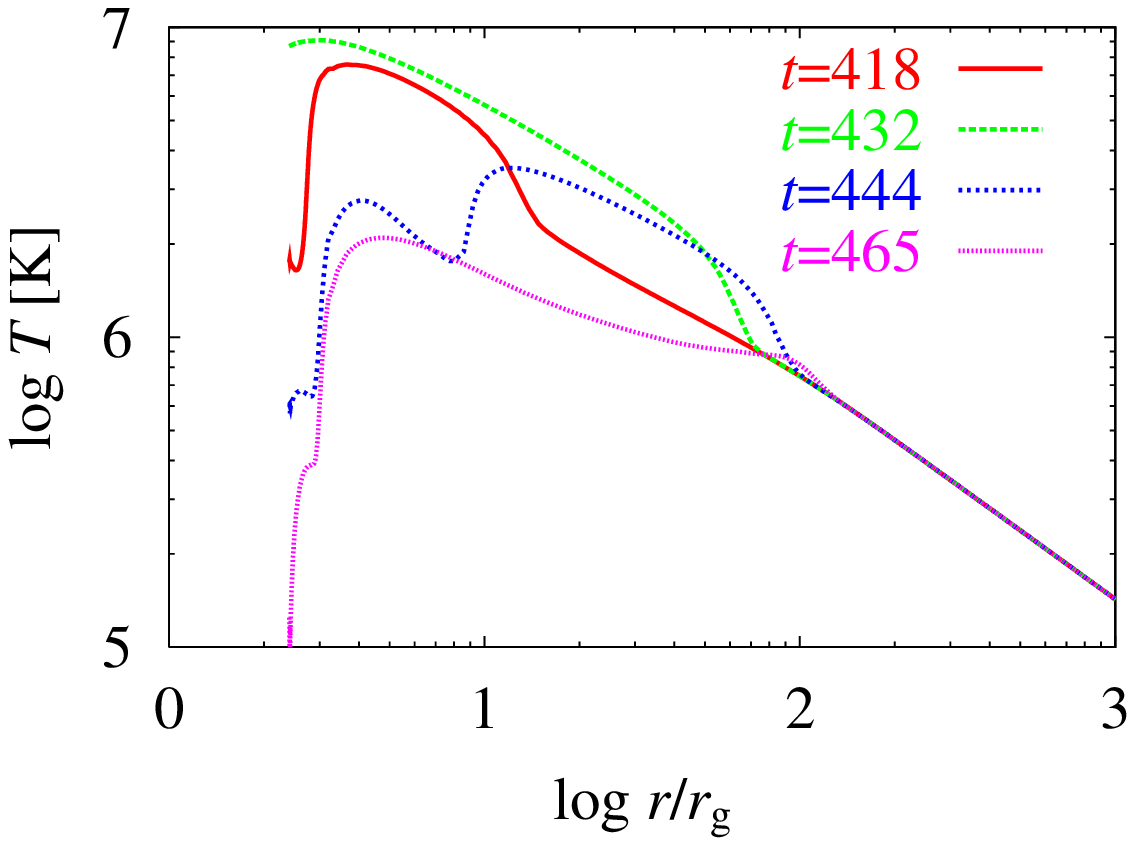}
\hspace{1.0cm}
    \FigureFile(75mm,75mm){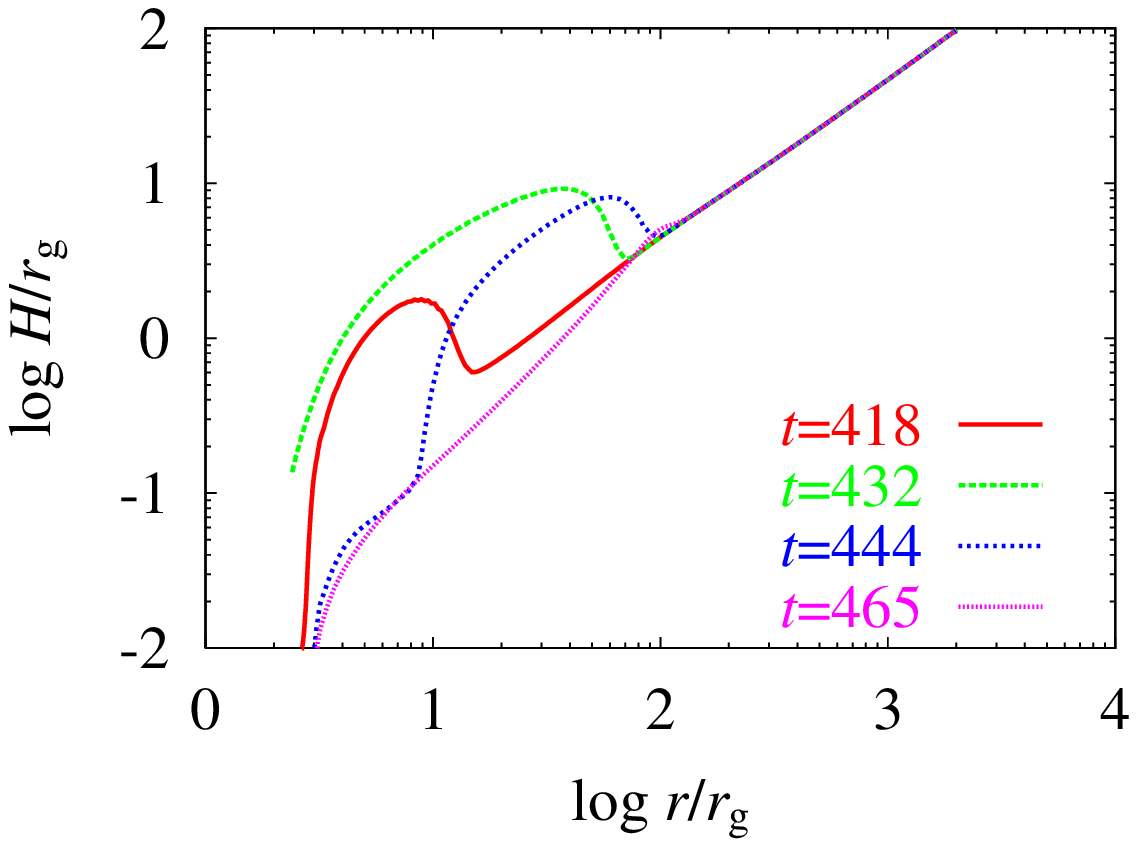}
}
  \end{center}
  \caption{Disk evolution during thermal instability;
the left and right panels represent 
the effective temperature profiles and the disk scale height, respectively.
 The sequences are before the thermal instability (thick solid line),
 at the peak (thick dashed line),
 just after (dotted line),
 30 seconds after thermal instability. 
}
  \label{fig1}
\end{figure}

As the accretion rate increases, 
 the radial velocity of the flow increases,
 and finally the emitted photons are trapped by the advective flow. 
That is, the advective cooling begins to work,
 $Q_{\rm adv}^- \propto \dot{M}^2$. 
After the mass falls into black hole, the disk cannot maintain the slim disk state,
thus the slim disk transits to the standard disk again. 
This transition is known as a ``limit-cycle oscillation''.
During the limit-cycle oscillation,
 the scale height and temperature distribution of the disk dramatically varies. 
We adopt the data by WM03 with the parameter set $\mu=0.1$, $\alpha=0.1$ and $m=10$.
Time evolution of the disk scale height and temperature distribution
 during the transition is shown in figure \ref{fig1} (WM03). 
Then, we verify that the geometrical thickness rapidly increases
 around the disk inner region ($\lesssim$ 100$r_{\rm g}$)
 during the disk transition due to the appearance of the radiation-pressure-dominant region.
The disk thickness is close to 45 degrees ($H / r \sim 1$) at the peak.
Therefore, we should pay special attention to
 the geometrical thickness of the disk. 

\section{Observed Characteristics during Thermal Instability}
We anticipate that photons emitted from an inner region are blocked by the increased scale height.
 In addition, photons from the inner region are affected by the special/general relativistic effects due to strong gravitational fields.
 Therefore, we expect that the image of an accretion disk is strongly affected by the scale height.
 The geometrical effect was considered in Madau (1988) and Fukue (2000).
 However, the special/general relativistic effects were not considered in these. The geometrical and relativistic effects have been considered in Watarai et al. (2005).
 Therefore, we pay special attention to the effect of scale height and make the images of the disk during thermal instability.
 To confirm how we see the accretion disk, we adopt the "Ray-Tracing Method'' (The details of this method is written in Luminet 1979; Fukue, Yokoyama 1988).

Photons emitted from any points on the disk travel along the null geodesics to be received by the distant observer.
 After the ray arrives at the disk surface,
 we obtain the physical quantities, such as,
the temperature, velocity field, and scale height,
 which are numerically-calculated disk data at the arriving point. 
 Due to relativistic effects, the observed radiative flux $F_{\rm obs}$ and blackbody temperature $T_{\rm obs}$ related to the emitted radiative flux $F_{\rm em}$ and temperature $T_{\rm em}$ as
\begin{eqnarray} 
F_{\rm obs} &=& \frac{F_{\rm em}}{(1+z)^4},
\label{eq:fobs} \\
T_{\rm obs} &=& 
  \frac{T_{\rm em}}{(1+z)},
\label{eq:tobs}
\end{eqnarray} 
using Lorentz invariance.
Here, redshift factor $z$ of the light emitted from a surface of the disk:
$1 + z = E_{\rm em} / E_{\rm obs} = L^{-1} \gamma D^{-1}$
 (see Kato et al. 1998),
where $L$ is the lapse function representing the gravitational redshift,
 $\gamma$ is the Lorenz factor denoting the transverse Doppler effect,
 and $D$ is the Doppler factor expressing the longitudinal Doppler effect.
\subsection{Imaging of the Disk during Thermal Instability}
The images of bursting disks and black hole shadows 
during thermal instability
 are presented in figure 2.
 \begin{figure}[hp]
  \begin{center}
\hbox{
    \FigureFile(180mm,160mm){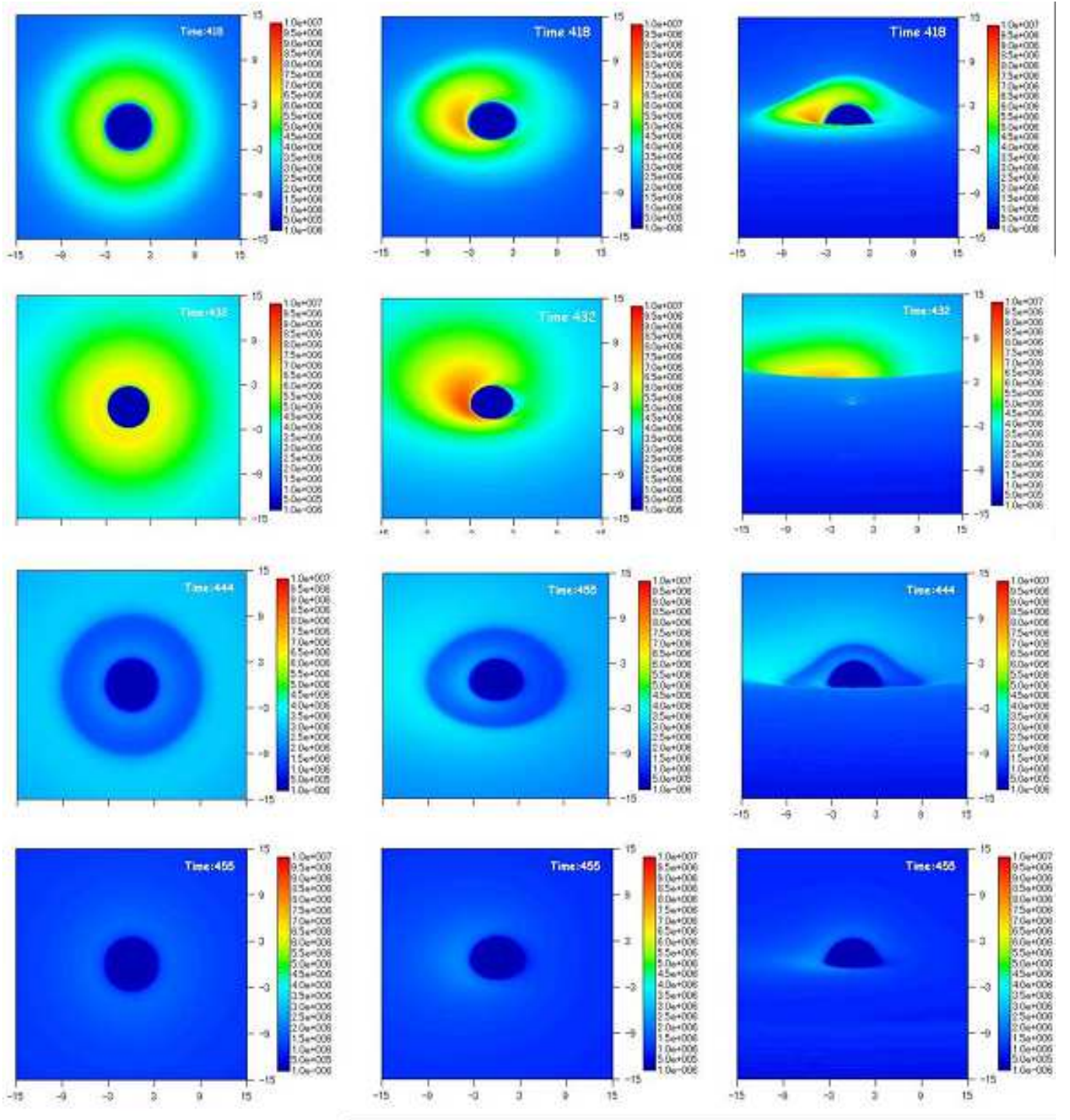}

}
  \end{center}
  \caption{Observed temperature distributions during thermal instability. 
  From top to bottom, before thermal instability ($t=418$ s),
 the peak of thermal instability ($t=432$ s),
 immediately after thermal instability ($t=444$ s),
 of 20 seconds after thermal instability ($t=455$ s).
Unit of the time-scale is second.
 From left to right;
 $i=1^{\circ}$, $i=50^{\circ}$ and $i=80^{\circ}$.
 The number of meshes is 250 $\times$ 250, and the screen size is
 15$r_{\rm g} \times 15 r_{\rm g}$.
Unit of X-Y axis is the Schwarzschild radius $r_{\rm g}$.
}
  \label{fig:obstemp}
\end{figure}
These figures represent the time evolution of the disk for 
several inclination angles 
($i=1^{\circ}$, $i=50^{\circ}$, and $i=80^{\circ}$ from left to right),
and of the states of the disk (before the burst to after the burst from top to bottom).

Let us see the image for $i=1^{\circ}$ (face on). 
As the mass-accretion rate increases, the effective temperature
 of the innermost region of the disk rises up to $\sim10^{7}$K, 
the high energy radiation is expected around the vicinity of the black hole. 
After the burst peak,
 the temperature in the disk innermost region descends rapidly.

 On the other hand, for an intermediate inclination angle ($i=50^{\circ}$), 
the emergent radiation from the left part, 
which approaches the observer with the relativistic speed near to the black hole is remarkably enhanced by the Doppler beaming effect (blue shift), 
while the right part is reduced (red shift). 
Moreover, the light ray originating from the opposite side of the disk is strongly bent by the gravity of the black hole before it reaches the observer. 
Therefore, the images of the accretion disk become asymmetric and distorted shapes. 
However, the effect of self-occultation is not remarkable in the case of $i=50^{\circ}$.
 The self-occultation effect remarkably appears in the case of a high inclination angle ($i=80^{\circ}$). 
The black hole shadow is completely hidden by the disk itself in the peak phase. This effect of geometrical thickness affects the luminosity and spectral energy distribution (SED) of the black hole system.
 We inspect this geometrical effect for the luminosity in the next subsection.

\subsection{Luminosity Evolution during Thermal Instability}
 \begin{figure}[htb]
  \begin{center}
\hbox{
    \FigureFile(70mm,70mm){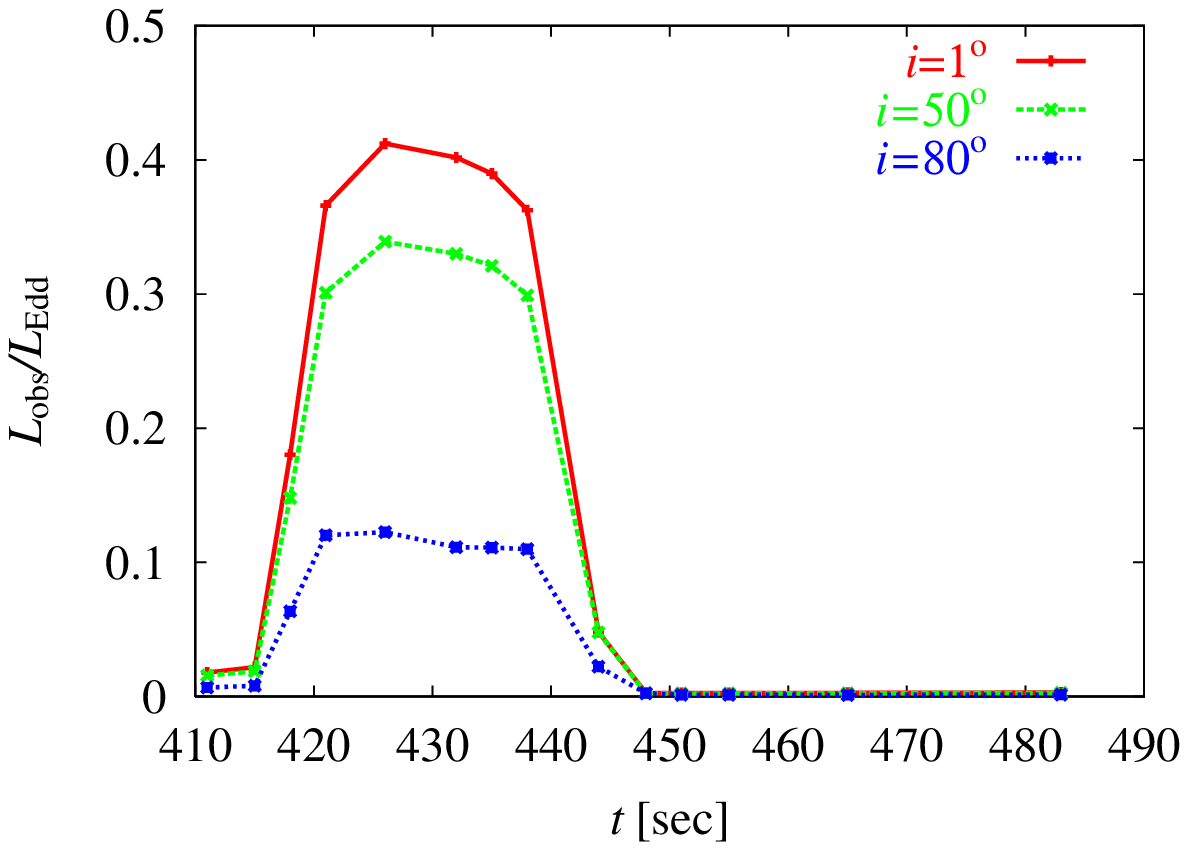}
}
  \end{center}
  \caption{Light curves during thermal instability for several inclination angles; $i=1^{\circ}$ (solid line), $i=50^{\circ}$ (dashed line), $i=80^{\circ}$ (dotted line). The abscissa is time in units of seconds, and the ordinate is the bolometric luminosity normalized by the Eddington one.
}
  \label{fig:effpot3}
\end{figure}
 Figure 3 represents the bolometric luminosity during thermal instability.
 The luminosity strongly depends on the viewing angle due to the projection effect and self-occultation. 
 When the viewing angle is $i=1^{\circ}$, only gravitational redshift dominates. 
Therefore, the luminosity of the disk varies 
from $L_{\rm obs}$ $\sim$ 0.01--0.4 $L_{\rm Edd}$. 
This result is consistent with that of WM03. 
As the inclination angle increases, the Doppler beaming and transverse redshift begin to be at work. 
However, the reduction of observed area mostly affects and observed luminosity decreases.
In the case of $i \gtrsim 50^{\circ}$, not only the effect of projection
 but also the self-occultation affect to the image of the disk.
 So the luminosity of the disk with $i=80^{\circ}$ decreases much more than $i=50^{\circ}$. 
Moreover, due to the self-occulting, 
the profile of the light curve for the case of $i=80^{\circ}$ is different from and flatter than the other cases of low inclination angles.

 To investigate how much the luminosity is blocked by the self-occultation effect, 
we plot the luminosity as a function of inclination angles (figure 4).
 \begin{figure}[htb]
  \begin{center}
\hbox{
    \FigureFile(70mm,70mm){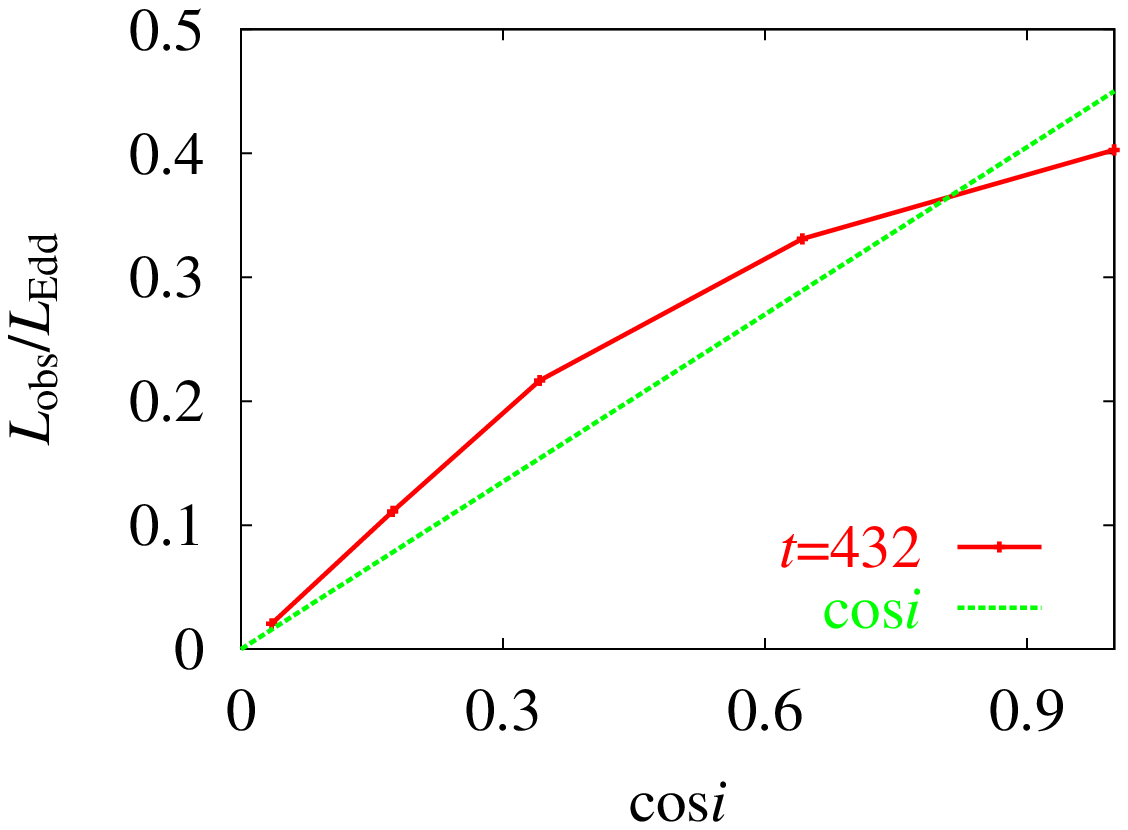}
}
  \end{center}
  \caption{Relationship between the peak luminosity and inclination angles $i$.
The abscissa is $\cos i$, and 
the ordinate is the disk luminosity normalized by the Eddington one. 
The dotted straight line is the projection factor 
under purely the geometrical effect. 
}
  \label{fig:effpot3}
\end{figure}
 In figure 4, we show the relation between the light curves of thermal instability at the peak and the influence of inclination angle.
 The case of cos $i$ $\gtrsim$ 0.8 falls below the evaluation considering only the projection effect, i.e., cos $i$ 
because the gravitational redshift dominates. 
In the case of 0.4 $\lesssim$ cos $i$ $\lesssim$ 0.8,
the disk luminosity exceeds the evaluation. 
This is because the Doppler beaming dominates compared with the gravitational redshift.
 However, in the case of cos $i$ $\lesssim$ 0.4, the geometrical effect of self-occultation dominants and the luminosity decreases.
Here, we emphasize that it is important to consider the geometrical effect for high inclination angles.

\subsection{Spectral Evolution during Thermal Instability}
We also calculated the spectra of the accretion disk. 
The spectra during thermal instability are affected by Comptonization
 due to their high temperature and large optical depth for electron scattering. 
Here, we adopt the diluted blackbody approximation as the effect of Comptonization
 for simplicity. The flux is given by 
\begin{equation}
F_{\nu}^{\rm db} = \frac{1}{f^4 (1+z)^3} \pi B_{\nu} (f T_{\rm eff})
\label{eq:p0}
\end{equation} 
where $\nu$, $B_\nu$, and $f$ are the frequency,
the Planck function and the spectral hardening factor
 (Ebisuzaki et al. 1984). 
The spectral hardening factor represents the ratio of the color temperature
 to the effective temperature. 
The actual value of $f$ is assumed to $f=1.7$
 which obtained from the radiation transfer calculation 
 in the vertical direction of the disk (Shimura, Takahara 1995). 
We note that our present equation (\ref{eq:p0}) includes 
the relativistic effect, i.e., $1/(1+z)^3$. 

The calculated spectral energy distributions of the accretion disk
 during thermal instability are presented 
(in actual, we plotted 
$ L_{\nu} = \int_{r_{\rm in}}^{r_{\rm out}} 2 F_{\nu}^{\rm db} \cdot 2 \pi r dr$)
 \begin{figure}[htb]
  \begin{center}
\hbox{
    \FigureFile(55mm,55mm){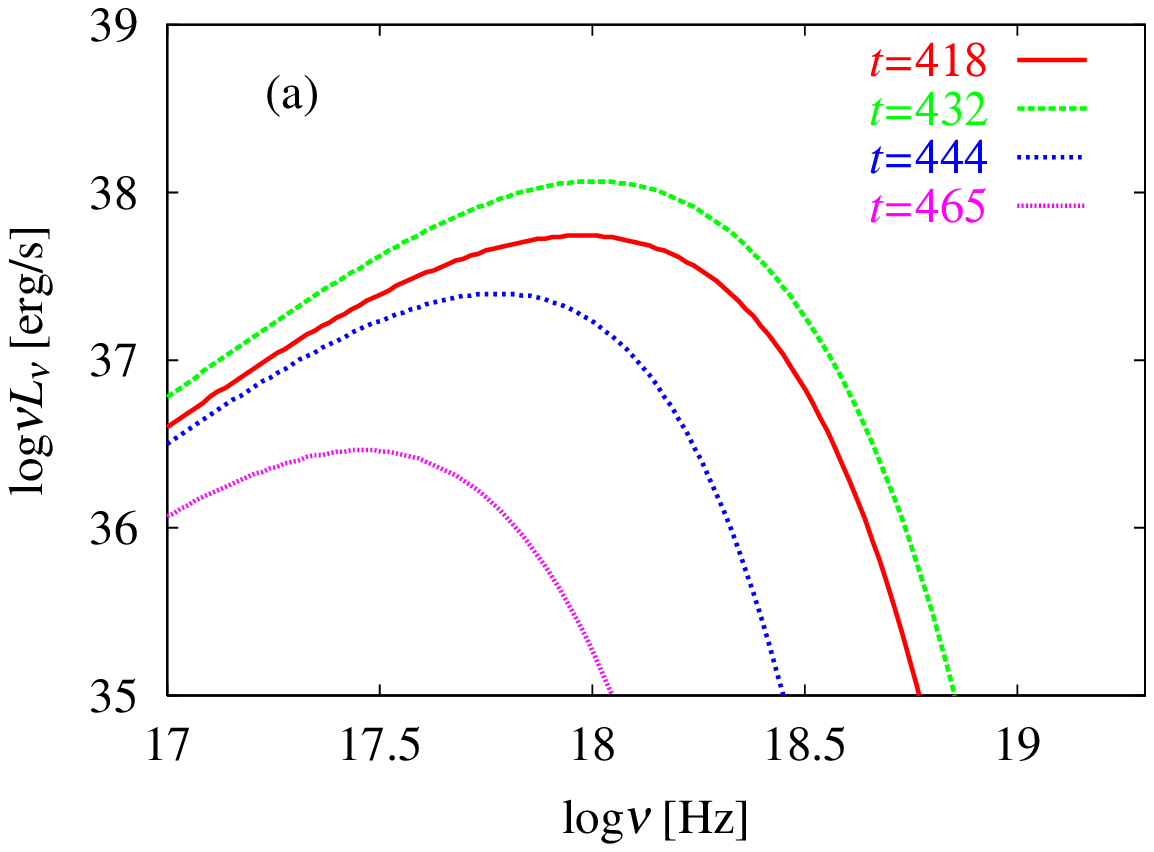}
\hspace{0cm}
    \FigureFile(55mm,55mm){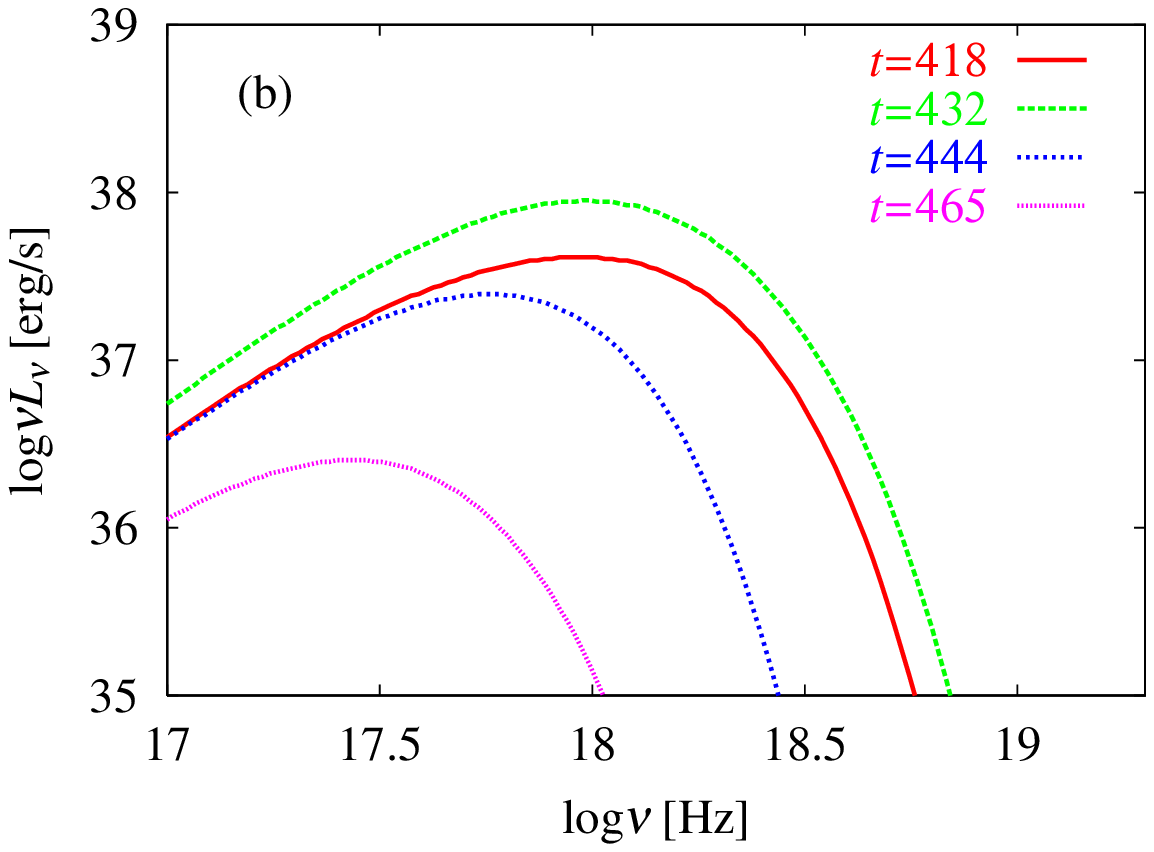}
\hspace{0cm}
    \FigureFile(55mm,55mm){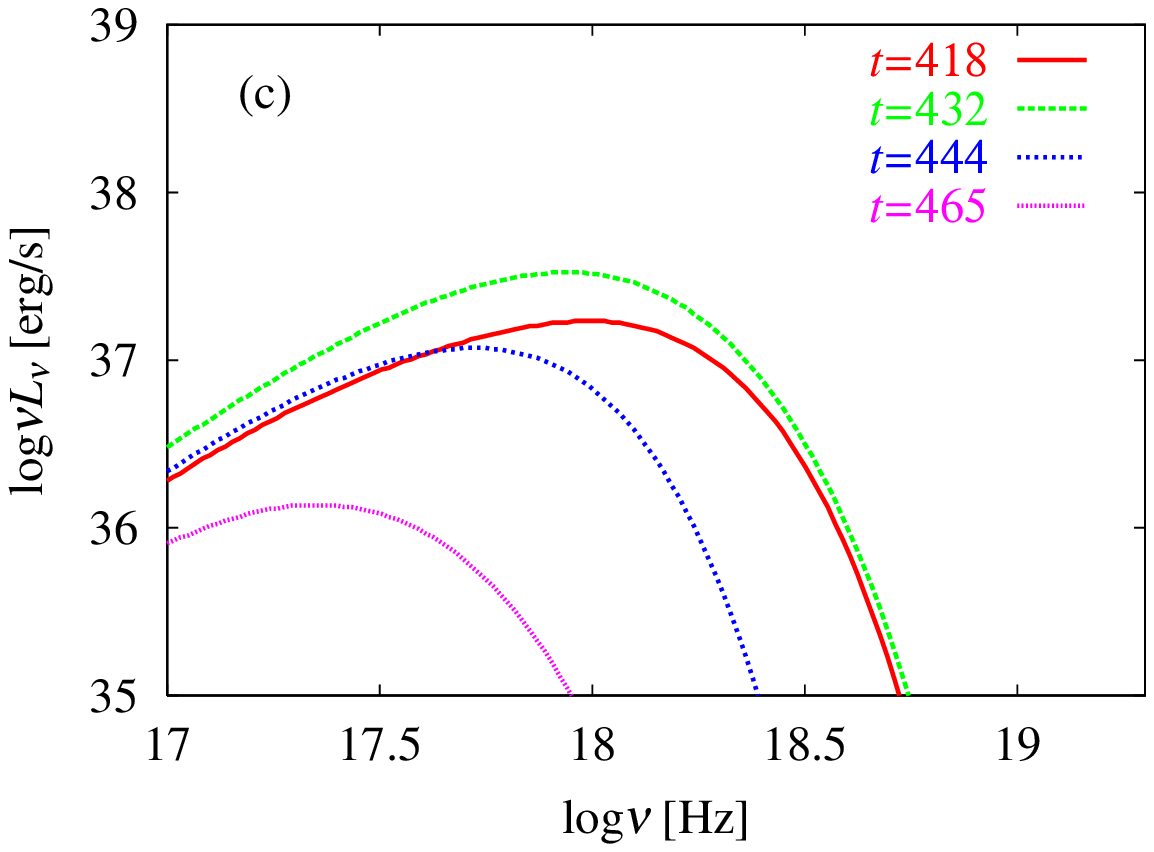}
}
  \end{center}
  \caption{Spectral energy distributions for several inclination angles with different time; before thermal instability (thick solid line), at the peak of thermal instability (thick dashed line), after thermal instability (dotted line), 30 seconds after finishing thermal instability. The inclination angle are (a) $i=1^{\circ}$, (b) $i=50^{\circ}$, and (c) $i=80^{\circ}$.
}
  \label{fig:spec}
\end{figure}
In figure \ref{fig:spec}, the spectrum shapes for $i=50^{\circ}$
 are similar to $i=1^{\circ}$ case throughout the time evolution. 
Because the geometrical thickness does not contribute to
 the image of the disk so much,
 the difference does not appear in the spectrum.
On the other hand, in the case of $i=80^{\circ}$,
 the self-occulted effect clearly appears in the SED. 
The high energy photons emitted from the innermost region are occulted by
 the large geometrical thickness,
 and the amount of the photons which finally reaching an infinite observer
 are decreases, and the spectrum becomes soft.
We therefore should pay attention to the inclination angle.

In this paper, we ignore the contribution of the coronal component.
If the high temperature corona expands in the vertical direction 
above the disk,
large amount of photons emitted from the disk are scattered by the electrons.
It is expected that the coronal component contributes to the high energy spectrum during thermal instability. 
Strictly speaking, we should solve the radiative transfer equation until the surface of the optical depth for electron scattering $\tau_{\rm es} \sim$ 1 surface including the coronal structure. 
However, we ignore such a coronal structure in this study
 in order to avoid a confusion of the origin of the spectrum.

\section{Discussion}
\subsection{Where Are Hidden Limit-cycle Busters?}
In the previous section, we have demonstrated that the observed flux
 from the inner region of luminous accretion disks
 strongly depends on the inclination angle.
According to our results, if other black hole candidates
 have a high inclination angle, it is difficult to detect
 the significant light curve variations. 
Because the disk outer rim blocks the emission from the disk inner region
 during the burst. 
This statement has already been pointed out by Watarai et al. (2005). 

Recently, type 2 counterparts of narrow line Seyfert 1 galaxies
 (hereafter NLS1s) have been
 discovered by several authors (Nagar et al. 2002; Dewangan, Griffiths 2005). 
Type 2 NLS1s candidates, NGC5506, NGC7582, and NGC7314, show  
 the signatures of rapid X-ray variability
 (a few $\times 10^2$ to $10^4$ seconds) and
 the reprocessed emission lines. 
According to the unified model of AGNs,
 type 2 AGNs has high inclination angles (Blandford 1990). 
That is, our present calculations can apply for 
 accretion disks in type 2 NLS1s.  
As for type 1 NLS1, supercritical accretion scenario 
 seems to be consistent with the observations
 because of their large luminosity
 even for small black hole mass (Mineshige et al. 2000; Collin 2002; Collin, Kawaguchi 2004). 
If we believe the supercritical accretion scenario, 
 we can predict that even if  type 2 NLS1s are really undergoing the thermal instability,
 the observed luminosity does not reach at the Eddington luminosity due to 
the self-occultation effect. 
Unfortunately, the decreasing of the luminosity is caused by
 a lot of reasons, 
 i.e., absorption of interstellar matters or molecular torus moving around AGNs. 
Thus, it is difficult to discriminate the cause of the decreasing of the luminosity. 
However, if it is possible to combine another physical information, such as
 time evolution of the reprocessed lines,  
 it provides us to constraint the origin of the absorbers or
 the geometry of the accretion disk.

\subsection{Observational Implications of Disk Line Spectrum}
Since the discovery of the broadened iron K$\alpha$ line at 6.4 keV in Seyfert galaxy MGC--6-30-15, 
 an iron K$\alpha$ line is widely used as a probe of the special/general relativistic effects
 near the supermassive black hole (Tanaka et al. 1995).  
The origin of line emission is roughly explained 
by the reflection model, i.e., 
the reflection component within the accretion disk or above the disk atmosphere
 absorbs the radiation from the intense X-ray source, 
and reradiates X-ray fluorescence lines (Miniutti, Fabian 2004). 
%
Actually, observed line spectra in some AGNs are consistent with above scenario.

While the spectrum of narrow line Seyfert 1 galaxy 1H0707-495
 shows more peculiar behaviors, for example, 
 the variability time scale is about factor of 4 shorter than the typical AGNs,
 the photon index is steeper than usual AGNs, 
 and the deep edge feature at about 7 keV (Boller et al. 2002; Tanaka et al. 2004).  
Fabian et al. (2002) assumed multiple reflection sheets in the radiation-pressure dominated disk
 to explain the deep edge at about 7 keV in NLS1 1H 0707-495. 
This object also shows a strong soft X-ray excess, and 
the observed disk temperature implies that the bolometric luminosity in 1H0707-495
 is close to the Eddington limit (Tanaka et al. 2004). 
When the disk is shining at the Eddington luminosity,
 the state of the disk should be in the state of slim disk rather than standard disk. 
If so, the reprocessed gas
 may be brown away by the strong radiation from the accretion disk,
 finally it may be difficult to reproduce the observed line intensity. 
We therefore should take into account the disk model more seriously   
 because the reprocessed emission is strongly influenced
 by the radiation field of the disk. 


The correlation between the continuum and line spectrum during thermal instability 
 seems to be important for investigations of the spectral features
 like the emission line or absorption edge. 
Because the emission from the disk inner region is blocked by the disk outer edge for high viewing angles. 
How much part is covered with the disk outer edge? 
How is the absorption edge made? 
How does the line flux correlate to continuum flux?
These questions remain as a question
 that we should be going to answer in the near future. 

\section{Conclusion}
We calculated the bolometric flux images and the observed spectra
 during thermal instability with considering the geometrical
 and special/general relativistic effects. 
Our results indicate that the observed images strongly
 depend on the viewing angles. 
In the case of low inclination angle ($i \lesssim 70^{\circ}$),
 the image of the disk does not change so much. 
Only Doppler boosting and special/general relativistic effects appear. 
So the distribution of luminosity and spectrum has the peak in the peak phase.

In contrast, if the inclination angle of the system
 is high ($i$ $\gtrsim$ $70^{\circ}$),
 the observed luminosity during the burst does change by much
 due to the self-occultation. 
Therefore, these facts indicate that if these situations occur
 in the real objects some bursting black holes still lurk in our Galaxy. 
If the jet from the accretion disk 
in the high luminosity state is related to the relaxation oscillation of the disk,
we could detect this bursting bevavior through the jet. 
The disk-jet interaction will be an intriguing issue in high luminosity objects
(Fender, Belloni 2004)

Moreover, if we predict the relationship between the continuum and the line during the disk evolution
 we may restrict the physical state of the object from the comparison with the observed data.
The evolution of the line spectra is one of our next issues.

\vspace{10mm}
We would like to thank Professors S. Mineshige and K. Seakale, and 
Dr. T. Kawaguchi
 for useful comment and discussions.
This work was supported in part by the Grants-in Aid of the
Ministry of Education, Culture, Sports, and Science and Technology 
(15540235, JF).
This research was partially supported by the 
Ministry of Education, Culture, Sports, and Science Technology Grants-in-Aid for JSPS Fellow (16004706, KW).

%
 

\begin{thebibliography}{}
\bibitem[]{}Belloni, T., Mendez, M., King, A. R., van der Klis, M., \& van Paradijs, J. 1997a, \apj, 479, 415
\bibitem[]{}Belloni, T., Mendez, M., King, A. R., van der Klis, M., \& van Paradijs, J. 1997b, \apj, 488, 109
\bibitem[]{}Belloni, T., Mendez, M., King, A. R., van der Klis, M., \& van Paradijs, J. 2000, \aap, 355, 271
\bibitem[]{}Blandford, R, D., Netzer, H., \& Woltjer, L. 1990, Active Galactic Nuclei (Springer-Verlag)
\bibitem[]{}Boller, Th., Fabian, A. C., Sunyaev, R., Tr$\ddot{\rm u}$mper, J., Vaughan, S., Ballantyne, D. R., Brandt, W. N., Keil, R., \& Iwasawa, K. 2002, \mnras, 329, 1
\bibitem[]{}Collin, S. Boisson, C. Mouchet, M. Dumont, A.-M. Coup\'{e}, S. Porquet, D. Rokaki, E. 2002, \aap, 388, 771
\bibitem[Collin \& Kawaguchi (2004)]{collin04} Collin, S.,
    \& Kawaguchi, T. 2004, \aap, 426, 797
\bibitem[]{}Dewangan, G. C., \& Griffiths, R. E. 2005, \apj, 625, 31
\bibitem[]{}Ebisuzaki, T., Sugimoto, D., \& Hanawa, T. 1984, \pasj, 36, 551
\bibitem[]{}Fabian, A. C., Ballantyne, D. R., Merloni, A., Vaughan, S., Iwasawa, K., \& Boller, Th. 2002, MNRAS, 331, 35
\bibitem[]{}Fender, A, P., \& Belloni, T, M. 2004, \mnras, 355, 1105
\bibitem[]{}Fukue, J., \& Yokoyama, T. 1988, \pasj, 40, 15
\bibitem[]{}Fukue, J. 2000, \pasj, 52, 613
\bibitem[]{}Honma, F., Matsumoto, R., \& Kato, S. 1991, \pasj, 43, 147
\bibitem[]{}H\=oshi, R. 1977, PTP, 58, 1191
\bibitem[]{}Kato, S., Fukue, J., \& Mineshige, S. 1998, Black-Hole Accretion Disks (Kyoto: Kyoto Univ. Press)
\bibitem[]{}Luminet, J.-P. 1979, \aap, 75, 235
\bibitem[]{}Matsumoto, R., Kato, S., Fukue, J. \& Okazaki, A. T. 1984, \pasj, 36, 71
\bibitem[]{}Madau, P. 1988, \apj, 327, 116
\bibitem[]{}Mineshige, S., Kawaguchi, T., Takeuchi, M. \& Hayashida, K. 2000, \pasj, 52, 499
\bibitem[]{}Miniutti, \& G. Fabian, A. C. 2004, MNRAS, 349, 1435
\bibitem[]{}Mirabel, I. F., \& Rodr\'{\i}guez, L, F. 1999, ARA\&A, 37, 409
\bibitem[]{}Nagar, N. M. Oliva, E. Marconi, A. \& Maiolino, R. 2002, \aap, 391, 21
\bibitem[]{}Paczy$\acute{\rm n}$sky, B., \& Wiita, P. J. 1980, \aap, 88, 23
\bibitem[]{}Shakura, N. I., \& Sunyaev, R. A. 1973, \aap, 88, 23, 432, 61
\bibitem[]{}Shakura, N. I., \& Sunyaev, R. A. 1976, MNRAS, 175, 613
\bibitem[]{}Shibazaki, N., H\=oshi, R. 1975, PTP, 54, 706
\bibitem[]{}Shimura, T., \& Takahara, F. 1995, \apj, 445, 780
\bibitem[]{}Szuszkiewicz, E., \& Miller, J, C. 1997, MNRAS, 287, 165
\bibitem[]{}Szuszkiewicz, E., \& Miller, J, C. 1998, MNRAS, 298, 888
\bibitem[]{}Tanaka, Y. Nandra, K. Fabian, A. C. Inoue, H. Otani, C. Dotani, T. Hayashida, K. Iwasawa, K. Kii, T. Kunieda, H. Makini, F. \& Matsuoka, M. 1995, Nature, 375, 659
\bibitem[]{}Tanaka, Y. Boller, T. Gallo, L. \& Keil, R. 2004, \pasj, 56, L9
\bibitem[]{}Teresi, V., Molteni, D., \& Toscano, E. 2004, \mnras, 351, 297
\bibitem[]{}Teresi, V., Molteni, D., \& Toscano, E. 2004, \mnras, 348, 361
\bibitem[]{}Watarai, K., \& Mineshige, S. 2003a, \apj, 596, 421
\bibitem[]{}Watarai, K., \& Mineshige, S. 2003b, \pasj, 55, 959
\bibitem[]{}Watarai, K., Ohsuga, K., Takahashi, R., \& Fukue, J. 2005, \pasj, 57, 513
\bibitem[]{}Zampieri, L., Nobili, L., Turolla, R., \& Belloni, T. 2001, \mnras, 72, 73
%
\end{thebibliography}
\end{document}